\documentclass[aps,prd,twocolumn,amsmath,superscriptaddress,amssymb,showpacs,floatfix,nofootinbib,longbibliography,preprintnumbers,longbibliography]{revtex4-1}
\pdfoutput=1
\usepackage{graphicx}
\usepackage{bm}
\usepackage{times}
\usepackage{slashed}
\usepackage{color}
\usepackage{aas_macros} 
\usepackage{slashed}
\usepackage{lipsum}
\usepackage{subfigure}
\usepackage{multirow} 
\usepackage{amsmath}
\usepackage{array} 
\usepackage{varwidth} 

\usepackage{hyperref}

\hypersetup{
     colorlinks   = true,
     citecolor    = magenta,
     urlcolor     = magenta,
     linkcolor    = magenta
}

\newcommand{\be}{\begin{equation}}
\newcommand{\ee}{\end{equation}}
\newcommand{\bea}{\begin{eqnarray}}
\newcommand{\eea}{\end{eqnarray}}

\begin{document}
\preprint{LTH-1335}

\title{Gamma-Ray Lines in 15 Years of Fermi-LAT Data: \\New Constraints on Higgs Portal Dark Matter}

\author{Pedro De La Torre Luque}
\thanks{{\scriptsize Email}: 
\href{mailto:pedro.delatorreluque@fysik.su.se}{pedro.delatorreluque@fysik.su.se},{\scriptsize ORCID}: \href{http://orcid.org/0000-0002-4150-2539}{0000-0002-4150-2539}}
\affiliation{Stockholm University and The Oskar Klein Centre for Cosmoparticle Physics,  Alba Nova, 10691 Stockholm, Sweden}
\affiliation{Polytechnic School, Universidad San Pablo-CEU, CEU Universities. 28668 Boadilla del Monte, Spain}
\author{Juri Smirnov}
\thanks{{\scriptsize Email}: \href{mailto:juri.smirnov@liverpool.ac.uk}{juri.smirnov@liverpool.ac.uk}; {\scriptsize ORCID}: \href{http://orcid.org/0000-0002-3082-0929}{0000-0002-3082-0929}}
\affiliation{Department of Mathematical Sciences, University of Liverpool,
Liverpool, L69 7ZL, United Kingdom}
\affiliation{Stockholm University and The Oskar Klein Centre for Cosmoparticle Physics,  Alba Nova, 10691 Stockholm, Sweden} 
\author{Tim Linden}
\thanks{{\scriptsize Email}: 
\href{mailto:linden@fysik.su.se}{linden@fysik.su.se},{\scriptsize ORCID}: \href{http://orcid.org/0000-0001-9888-0971}{0000-0001-9888-0971}}
\affiliation{Stockholm University and The Oskar Klein Centre for Cosmoparticle Physics,  Alba Nova, 10691 Stockholm, Sweden} 

\begin{abstract}
Monoenergetic $\gamma$-ray spectral lines are among the cleanest signatures of dark matter annihilation. We analyze 15 years of Fermi-LAT data, find no spectral lines, and place strong constraints on dark matter annihilation to monoenergetic $\gamma$-rays. Additionally, we produce the first double-line analysis of the coupled signals from $\gamma\gamma$ and $Z \gamma$ lines, which proves particularly powerful for dark matter masses above $\sim150$~GeV. From our constraints on a double-line feature, we investigate and constrain some minimal models where the Galactic Center Excess (GCE) can be fit by dark matter annihilation through the Higgs boson into Standard Model particles. 
\end{abstract}

\maketitle

\section{Introduction}

Models that generate the observed dark matter (DM) abundance through thermal freezeout provide one of the most compelling explanations for the cosmological evolution of our universe~\cite{Zeldovich:1965gev,Lee:1977ua,Steigman:1984ac,DEramo:2010keq,Hochberg:2014dra,Hochberg:2014kqa,Farina:2016llk,Bertone:2016nfn,Dey:2016qgf,Cline:2017tka,Dolgov:2017ujf,Arcadi:2017kky,Roszkowski:2017nbc,Dey:2018yjt,Maity:2019vbo,Kim:2019udq,Smirnov:2020zwf,Asadi:2021bxp,Parikh:2023qtk}. Fortunately, scenarios dominated by a $2 \rightarrow 2$ process (e.g. $\chi\chi \rightarrow \text{SM SM}$) also provide us with a precise, testable target for DM annihilation searches~\cite{Steigman:2012nb,Bringmann:2020mgx}.
The thermal freeze-out mechanism does not generically predict the Standard Model (SM) final states or branching ratios, and thus it is common to examine DM models dominated by tree-level annihilations to different standard-model particle states, such as $b\bar{b}$, $\tau^+\tau^-$, $W^+W^-$ or other leptonic and hadronic pairs. 

However, these models are simplified for two reasons. First, DM annihilation may include tree-level couplings to a number of final states, with branching ratios that depend on the decay widths of the intermediate particles. Second, in addition to tree-level processes, there are guaranteed loop-level processes. Some of these final states, like those that produce $\gamma\gamma$ or Z$\gamma$ lines, may be more detectable than tree-level annihilation processes despite their subdominant branching fractions.

While the two-photon channel leads to a mono-energetic line at $E_\gamma = m_{\rm DM}$. The $Z\gamma$ channel is kinematically accessible at DM masses of $m_{\rm DM} > m_Z/2$, and leads to final state photons 
with energies centered around
\vspace{-0.2cm}
\begin{equation}
    E_{Z \gamma} = E_{\gamma \gamma} \left(1 - \left(\frac{m_{Z}}{2E_{\gamma \gamma}}\right)^2\right)\,.
    \label{eq:Energies}
\end{equation}
The spectrum has an intrinsic width due to the finite life-time of the $Z$ boson, which is given by~\cite{Beenakker:1988pv,Abrams:1989yk}
\begin{figure}[th!]
\includegraphics[width=0.95\columnwidth]{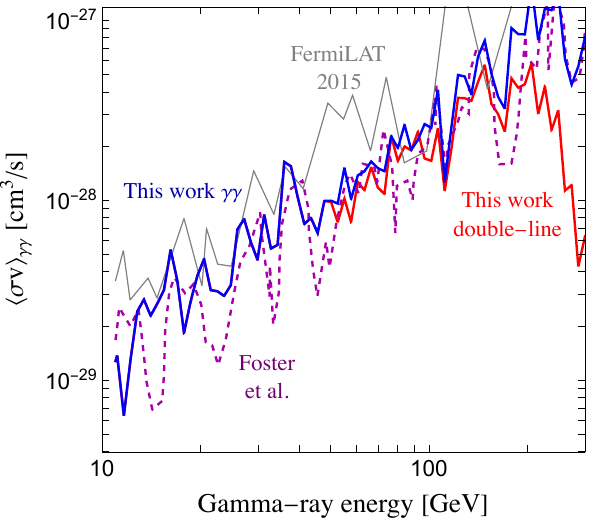}
\vspace{-0.4cm}
\caption{The results of our model-independent single-line (blue solid) and Higgs-portal double-line (red solid) analysis, given an NFW profile in ROI41. Single-line analyses by the Fermi-LAT collaboration (2015)~\cite{Fermi2015}(gray solid) and a recent result by Ref.~\cite{Foster:2022nva} (magenta dashed) are shown for comparison.}\vspace{-3mm}
\label{fig:LinesCompare}
\end{figure}
\vspace{-0.2cm}
\begin{align}
    \Gamma_{Z \gamma} \approx \frac{\Gamma_Z m_{Z}}{2 \sqrt{3} E_{\gamma \gamma} } < \frac{\Gamma_Z}{\sqrt{3}} \approx 1.2 \text{ GeV}\,.
\label{eq:Energies2}    
\end{align}
This width leads to an effectively monochromatic signal at the current Fermi-LAT energy resolution. Making use of this fact, we propose a double-line search that further increases our experimental sensitivity.

The relationship between the branching ratios to all final states depends on the mediator choice.  The simplest, and most predictive scenario is DM coupling though the Higgs portal, a singlet operator $H^\dagger H$~
\cite{Silveira:1985rk,McDonald:1993ex,Burgess:2000yq,OConnell:2006rsp,Cline:2013gha,Duerr:2015mva,Duerr:2015aka,Duerr:2015bea}. In this case, the branching ratios to all SM final states are entirely fixed by the well-known properties of the Higgs boson. Thus, combining collider-grade accuracy with the freeze-out condition entirely fixes our signal expectation for a given DM mass.

In this work, we reanalyse existing Fermi-LAT data, choosing CLEAN events from 180 months of the PASS8 data~\citep{Atwood2009apj}, and perform both single- and double-line searches for annihilating DM. We find that double-line analyses have a superior constraining power, especially at large DM masses. Furthermore, we show that our limits on Higgs-mediated DM annihilation are in tension with Higgs-portal interpretations of the GCE at masses near the $m_H/2$ resonance~\cite{Duerr:2015bea,Fraser:2020dpy}.

Fig.~\ref{fig:LinesCompare} shows the limits of our single- and double-line analyses on the $\langle \sigma v\rangle_{\gamma \gamma}$ annihilation cross section, and compares our results with previous work~\cite{Fermi2015, Foster:2022nva}. Our analysis provides stronger constraints, particularly at large $\gamma$-ray energies.

\section{Methodology and analysis}
\label{sec:Method}

\subsection{Gamma-ray datasets}
\label{dataset}
The Fermi-LAT is a pair-conversion telescope that measures $\gamma$-rays with energies between $\sim$20~MeV and $\sim$1 TeV~\citep{Atwood2009apj}.
In this paper, we use $\sim$180 months of data spanning from 2008-08-04 to 2023-07-20 selecting CLEAN events from the PASS8 data. We include events from good quality time intervals and remove periods when the LAT was operating at rocking angles $\theta_r > 52\deg$ ((DATA\_QUAL$> 0$) \&\& (LAT\_CONFIG==1) \&\& ABS(ROCK\_ANGLE)$<52$). We also apply the zenith-angle cut $\theta_z < 90\deg$, to avoid contamination from the Earth limb. We limit our analysis to EDISP3 events \textit{evtype $= 512$}), which have the best energy reconstruction (hence, best energy resolution), to minimize uncertainties relating to instrumental energy dispersion. We employ the P8R3\_CLEAN\_V3 version of the instrument response functions. The extraction of Fermi-LAT data and calculation of exposure maps is performed using the most up-to-date version of the ScienceTools\cite{Fermitools} (2.0.8; released on 01/20/2021). In addition, we performed different consistency checks of our analyses using PASS7 and PASS8 front- and back-converted events, which allows us to compare our results with those obtained by the Fermi-LAT Collaboration~\cite{Fermi2013}.

We extract data spanning from $10$ GeV to $300$ GeV in $130$ logarithmically uniform energy bins, which constitute an energy resolution $\Delta E/E \sim 3\%$ (which is better by at least a factor of two than the intrinsic energy resolution of EDISP3 events). Closely following previous Fermi analyses, we divide our data into three regions of interest (ROIs), which are spherical regions of $3$, $16$, and $41$ degrees around the Galactic center, focusing on the inner regions studied by the Fermi collaboration in Refs.~\cite{Fermi2013, Fermi2015} (see also~\cite{Weniger2012} for a similar approach). For each ROI, the regions of $68\%$ containment around known sources are subtracted in every dataset following the \textit{4FGL\_DR2} catalog, except for the ROI3 region, where no source subtraction is done. We also mask the galactic plane in regions with $|b|<5^{\circ}$ and $|l|>6^{\circ}$ as in Refs.~\cite{Fermi2013, Fermi2015}.

\subsection{Single and double line analyses}
\label{sec:Analysis}
We closely follow the strategy used by the Fermi collaboration in Ref.~\cite{Fermi2013} and search for spectral lines by performing maximum likelihood fits in each of our ROIs and in $88$ sliding energy intervals (the sliding window technique from Ref.~\cite{Gilles_Vertongen_2011}) from $10$~GeV to $300$~GeV. For each interval, we fit the count spectrum in an energy window that surrounds the central energy with a width of $\Delta E \geq 3\times \sigma_E(E)$\footnote{We repeated the analysis for window sizes between $2$ and $6\times \sigma_E(E)$ and found that the chosen width has almost no effect on our results.} (where $\sigma_E(E)$ is the half width of the $68\%$ exposure-weighted energy resolution of each dataset (See~\cite{Performance}), assuming that the background is described by a power-law and adding a line-like signal of free amplitude, which is smeared due to the energy resolution of the Fermi-LAT (Following~\cite{EDISP}). 
We use a likelihood function described by a Poisson distribution in the number of events at every energy window:
\begin{equation}
    \mathcal{L}^{ROI} = \prod_i \frac{ e^{-n(E_i, E'_i)} \times n(E_i, E'_i)^{N_i(E_i)} }{N(E_i)!}, 
    \label{eq:Likelihood}
\end{equation}
where $N_i$ is the observed number of events at energy $E_i$ in each dataset (ROI), and $n_i$ is the expected number of counts at energy $E'$ that are reconstructed at energy $E$ by the instrument. Under the null hypothesis, there is no line-like signal and the expected number of counts is found by fitting the data to a power-law describing the background emission ($n_i = n_{bkg}$). In the alternative hypothesis, the number of counts is described as $n_i = n_{sig}D_{eff} + n_{bkg}$, where $D_{eff}$ is the energy dispersion matrix that allow us to account for the energy reconstruction of the events by the LAT, and which was obtained using the \textit{gtdrm} Fermitool. 

It is important to remark that, within the energy range where we perform this analysis, the systematic uncertainties in the spectral reconstruction are expected to be negligible compared to the statistical uncertainties in the photon count, as reported by Refs.~\cite{Fermi2013, Fermi2015}. This assumption does not hold at much lower energies, which would require more complex modeling of Fermi-LAT responses, as in Ref.~\cite{Fermi2015}. The inclusion of systematic uncertainties would only slightly weaken our bounds at low energies, leaving the main conclusions of this manuscript unchanged.

To perform the fits we rely on the Markov Chain Monte Carlo (MCMC) package \textit{Emcee}~\cite{Emcee}, since this technique is more robust than conventional optimizers and less prone to finding false local minima. This analysis produces probability distribution functions for every parameter in the fit, which are used to estimate the credible intervals of each parameter and the DM limits. Since the best-fit number of signal events that we obtain is very low, we use the Feldman-Cousins (FC) method~\cite{FC}\footnote{We acknowledge the use of the package from \url{https://github.com/usnistgov/FCpy/tree/main}.} to ensure that we are not mis-evaluating the confidence intervals and, hence, the limits. Using the FC method produces roughly the same upper-limits as the MCMC algorithm, except when the best-fit number of source counts is very small. Concretely, we take the best-fit values for the number of background events and number of signal events obtained from the MCMC procedure, and apply the FC method 
with the likelihood functions defined in Eq.~\ref{eq:Likelihood}.
In this way, we reject unrealistically strong upper-limits, particularly for the downward fluctuations.

For the \textit{double-line analysis}, we repeat the same procedure as the single-line analysis, but add a correlated second line signal that accounts for DM annihilation through the Higgs into a second $Z\gamma$ line ($\chi + \chi \rightarrow H \rightarrow Z + \gamma$). 
The relative amplitude of the $\gamma \gamma$ and Z$\gamma$ signals is correlated by the branching ratio of the Higgs boson to each channel. 
Moreover, the energy of the line signal for $Z \gamma$ production is connected to the energy of the $\gamma\gamma$ signal as described by Eq.~\ref{eq:Energies}, and we set $E_{\gamma \gamma} = m_{\chi}$ for the annihilation process that we are considering. 

To account for the fact that the energy window in our double line analysis must accommodate both the $\gamma \gamma$ and Z$\gamma$ line energies, the lower edge of the energy window considered in this analysis is set to be the minimum value between $E_{Z\gamma}$ and $E_{\gamma \gamma} - 3\sigma_E$, to cover the double-line feature. This results in a larger energy window for the double-line analysis from $50$ to $\sim80$~GeV, above which the lower limit of the sliding energy window coincides with the one used in the single-line analysis (i.e. the lower limit is always the $3\sigma_E$ above $80$~GeV).

\section{DM bounds from the line search}
\label{sec:Results}
The expected $\gamma$-ray flux from the annihilation of DM particle through the process $\chi + \chi \rightarrow \gamma + \gamma$ in a region of the sky with angular size $\Delta\Omega$ is
\begin{equation}
    \frac{d\Phi}{dE} = \frac{1}{8\pi} \frac{\left< \sigma v \right>_{\gamma \gamma}}{m^2_{\chi}} \left(\frac{dN}{dE}\right)_{\gamma \gamma} \times \mathcal{J^{ROI}}(\Delta \Omega)  \,\,\,\,\, ,
\end{equation}
where $\mathcal{J^{ROI}}(\Delta \Omega)$ is the astrophysical J-factor that describes the expected annihilation rate given a specific choice of ROI and a DM distribution, $\left(\frac{dN}{dE}\right)_{\gamma \gamma} = 2\times \delta(E - E_{\gamma \gamma})$ is the $\gamma$-ray yield per annihilation, $m_{\chi} = E_{\gamma \gamma}$ is the mass of the WIMP, and $\left< \sigma v \right>_{\gamma \gamma}$ is the annihilation rate to the $\gamma \gamma$ channel and is related to the total annihilation rate via the mediator-dependent branching fraction $\left< \sigma v \right>_{\gamma \gamma} = \text{BR}_{\gamma \gamma} \times \left< \sigma v \right>_{\rm ann}$, where $\left< \sigma v \right>_{\rm ann}$ is the total DM annihilation rate.

For the $Z\gamma$ process, this same formula holds, but with $\left< \sigma v \right>_{Z \gamma} = \text{BR}_{Z \gamma} \times \left< \sigma v \right>_{\rm ann} $ and $\left(\frac{dN}{dE}\right)_{Z \gamma} = \delta(E - E_{Z \gamma})$, where the photon is produced at the energy given by Eq.~\ref{eq:Energies}.
The main uncertainty in deriving limits on the annihilation rate is the J-factor,  $\mathcal{J^{ROI}}(\Delta \Omega)$, which directly depends on the Milky Way DM distribution. Here, we assume a local DM density of $0.4$~GeV~cm$^{-3}$~\cite{Fabio_Iocco_2011} and a distance from the Solar System to the GC of $8.5$~kpc. We characterize two DM distributions, the NFW and a contracted-NFW profile with an index $\gamma = 1.3$ (motivated by studies of the GCE~\cite{FermiGCE_2017, DiMauro_GCE}), both with a scale radius of $r_s = 20$~kpc.

\subsection{Significance for lines in the gamma-ray spectrum}
\label{sec:TS}

\begin{figure}[!t]
\centering
\includegraphics[width=\columnwidth]{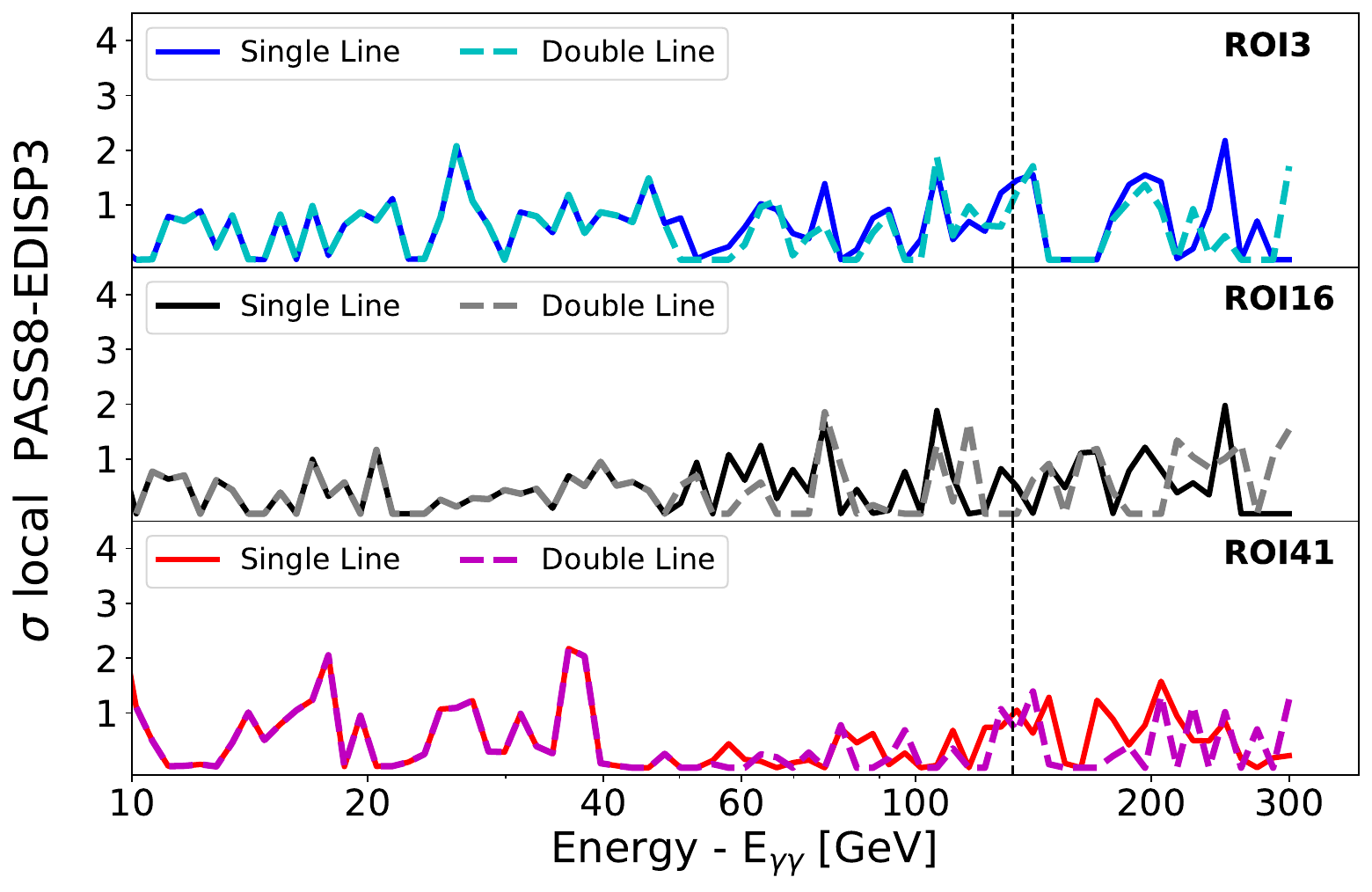} 
\caption{Local significance obtained in the $10$ to $300$~GeV range. We compare the result of the single-line analysis (solid lines) and double-line analysis (dashed lines) for each ROI: $3^\circ$  (upper), $16^\circ$ (middle) and $41^\circ$ (lower) around the center of the Galaxy. For those signals where there is no preference for a positive number of counts over the background the significance ($\sigma_{local}$) is set to 0.}
\label{fig:TS_comp}
\end{figure}

Figure~\ref{fig:TS_comp} shows the test-statistic (TS) computed for the single-line and double-line analyses as a function of the DM mass. This produces an accurate calculation (assuming Wilks' theorem holds) for the local significance of any line signal ($\sigma_{local} \sim \frac{n_{line}}{\sqrt{n_{bck}}}$), which can be calculated as
\begin{equation}
    {\rm TS} = 2\frac{\mathcal{L}(n_{sig} = n_{sig, {\rm Best}} )}{\mathcal{L}(n_{sig} =0)} .
\end{equation}
Although, the J-factor constitutes the largest uncertainty on the expected annihilation signal from the GC region, we remind the reader that the TS is independent of the J-factor employed. 
We note no statistically significant peaks (exceeding a 3$\sigma$ local significance) in any dataset. The most statistically significant peaks hardly exceed $2\sigma$ and are not repeatedly present in the different ROIs (i.e. a fluctuation not present in all the datasets).
We have performed an analogous analysis, without fixing the branching ratio between the $Z\gamma$ and $\gamma \gamma$ processes to the values predicted by the Higgs portal. Also in this general case no significant excess signals were observed and the local significance is roughly identical to the one obtained in the double-line analysis.

\subsection{DM bounds from the single-line and double-line analyses}
\label{sec:Bounds_SL}

Given the lack of significant excesses in the $\gamma$-ray spectrum, we derive the confidence limits for both the single- and double-line analyses. We produced the bands depicting the $68\%$ and $95\%$ confidence intervals of the observed limits by generating mock data following a power-law distribution with spectral index of $-2$ and Poissonian noise. We repeat the analysis for $750$ iterations (we find that the bands remain stable above $\sim400$ iterations), similar to the approach of Refs.~\cite{Fermi2013, Fermi2015}.

\begin{figure}[!t]
\centering
\includegraphics[width=\columnwidth] {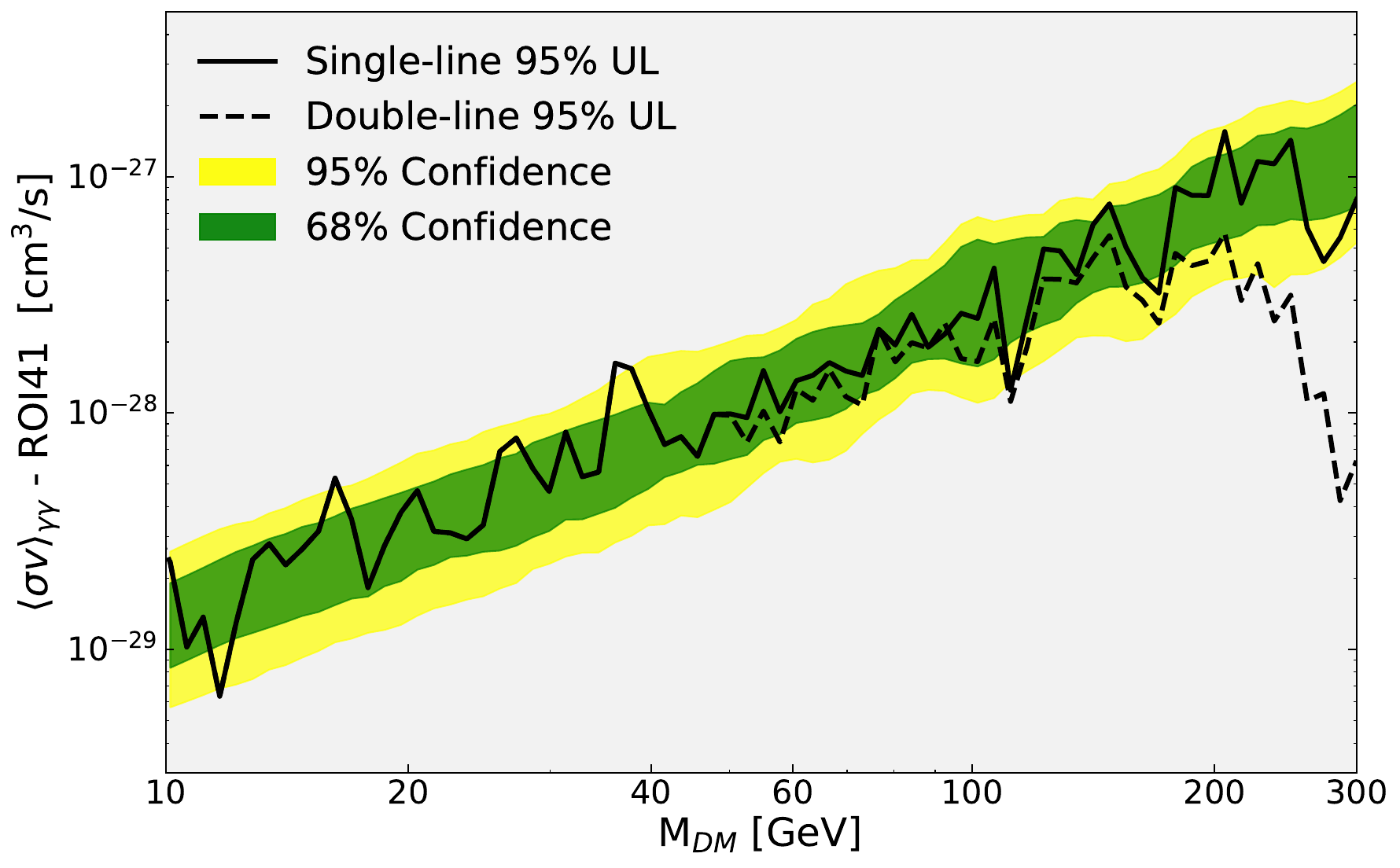}
\caption{DM bounds for the ROI41 region assuming a NFW Galactic DM profile, including the $68\%$ and $95\%$ confidence intervals obtained from the single-line analysis. We show the single-line and double-line limits as a solid and a dashed line, respectively, and compare to the limits from Refs.~\cite{Fermi2015}.}
\label{fig:DM_limits}
\end{figure}

Figure~\ref{fig:DM_limits} shows the derived limits for ROI41, which provide slightly stronger (but similar) limits than the other ROIs. We show results for other ROIs in the Figure~\ref{fig:DM_limits_SL} in Appendix~\ref{sec:AppendixB}. We also include the observed limits for the double-line analysis as a dashed line, finding that these constraints become stronger at higher energies when the $Z\gamma$ cross-section becomes larger than the $\gamma \gamma$ cross section.  The ratio BR$_{Z\gamma}$/BR$_{\gamma \gamma}$ is derived from~\cite{LHCHiggsCrossSectionWorkingGroup:2011wcg} and is discussed in the supplemental material.

In Figure~\ref{fig:LinesCompare}, we compare our single-line limits with those obtained by the Fermi-LAT collaboration (using $5.8$~yr of data)~\cite{Fermi2015} and those obtained by Foster et al.~\cite{Foster} (using $\sim$14~years of Fermi-LAT data). The main differences between our analysis and Ref.~\cite{Foster} include: (1) their analysis employs a constant energy-window size (of $\Delta E/E\sim 0.64$), while our analysis utilizes a variable window size based on the local Fermi-LAT energy resolution, (2) their analysis employs a single ROI focused on the inner $30^{\circ}$ around the GC and utilizes a different Galactic plane cut, (3) they utilize SOURCE class photon events with energy reconstructions spanning EDISP 1--3 (the top 75\% of well-reconstructed energy events) while we use only CLEAN class photon events from EDISP3 (the $25\%$ of reconstructed events with the best energy resolution) (4) they do not subtract regions surrounding bright $\gamma$-ray point sources, while we eliminate these background-dominated regions from our analysis in all ROIs except for ROI3. Despite these differences, our single-line results are in good agreement, as seen in Fig.~\ref{fig:LinesCompare}. 
The DAMPE collaboration recently published the results of their single-line analysis using $5$~yr of data collected by the DAMPE instrument~\cite{DAMPE_Lines}, obtaining similar bounds to those found in Ref.~\cite{Fermi2015} (see also Ref.~\cite{cheng2023search}).

\section{Implication for Higgs-Mediated Annihilation}

Higgs mediated annihilation is unique from the perspective that collider level precision can be used in a DM framework. Since SM processes govern the branching ratios for the annihilation final states, and the DM coupling to the Higgs is fixed by the freeze-out condition, the only unknown parameter is the DM mass.

However, there are several model choices that affect the relationship between the annihilation cross-section and the expected event rates in direct detection and collider experiments~\cite{Fraser:2020dpy,Abdallah:2015ter}. Here we mention two well-motivated models:

\begin{itemize}
    \item A singlet scalar model $S$, with mass $m_S$, which after the electroweak symmetry breaking has the following relevant coupling to the Higgs boson, $h$
\begin{align}
\label{eq:ModelScalar}
    -\mathcal{L} \supset  \frac{1}{2} m_S^2 S^2 + \frac{\lambda_p v_H}{2} h S^2\, , 
\end{align}
where $\lambda_p$ is a dimensionless coupling, and $v_H$ is the Higgs field vacuum expectation value~\cite{Silveira:1985rk,McDonald:1993ex,Burgess:2000yq}. In this case the spin-independent direct detection cross section is not suppressed, and only $\lambda_p < 10^{-3}$ values are compatible with the current limits~\cite{Duerr:2015aka,LZ:2022ufs}. The GCE signal can thus only be explained in a very narrow mass range around the Higgs resonance~\cite{Duerr:2015bea}. 
\item A Majorana fermion model $\chi$, with mass $m_\chi$ and the following low-energy couplings
\begin{align}
\label{eq:ModelFermion}
    -\mathcal{L} \supset  \frac{1}{2} m_\chi \bar{\chi} \chi + i \frac{y_p}{2} h \bar{\chi} \gamma_5 \chi\, 
\end{align}
where $y_p$ is a dimensionless coupling parameter. In this case the direct detection cross section is suppressed for real values of $y_p$~\cite{Fraser:2020dpy,Foster:2022nva}, which means that we would not expect a signal in direct detection searches. Therefore, this model is not constrained by direct detection experiments, and only visible in collider and indirect detection searches. 
\end{itemize}

We note that both scenarios are subject to invisible Higgs decay constraints, which limit $\text{BR}(h \rightarrow \text{inv.}) < 0.11$~\cite{Wang:2841498}, and difficult to avoid.

The total annihilation cross section in both scenarios is given by 
\begin{align}
 \sigma_{\rm ann.}(s) = \frac{f(m_{\rm DM}, \lambda_{\rm DM}) }{ \left( s -m_h^2 + \Gamma_h^2 m_h^2 \right)^2 }\,
\end{align}
where $s$ is the total s-channel four-momentum, $m_{\rm DM}$ and $\lambda_{\rm DM}$ are the DM mass and coupling strength to the Higgs, and $\Gamma_h$ is the total Higgs boson decay width. As discussed in Ref.~\cite{Gondolo:1990dk} the thermally averaged cross-section at a resonance needs to be performed without the non-relativistic expansion of the annihilation cross section. The fact that the thermal average of the annihilation cross section in the early universe is significantly different from the average at late times leads to a strong late-time mass dependence of the annihilation cross section around the resonance, as discussed in detail in Ref.~\cite{Duerr:2015aka}.  

\begin{figure}[t!]
\includegraphics[width=\columnwidth]{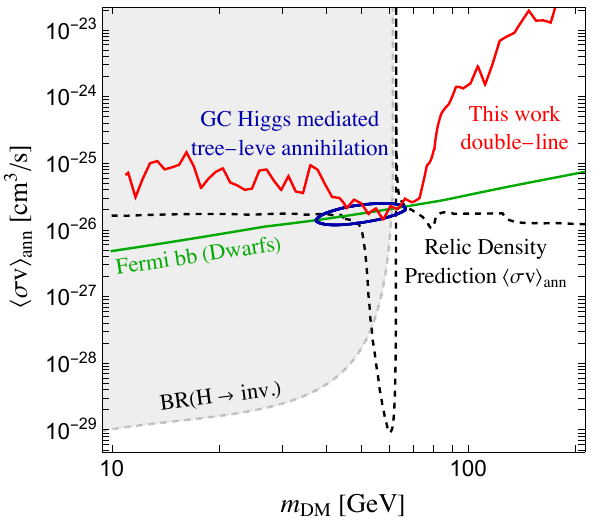}
\caption{ The Higgs portal total annihilation rate as a function of the DM mass. We superimpose our constraint from the double-line analysis on the total annihilation rate (red solid). Additionally we show the dwarf spheroidal limits from the Fermi-LAT collaboration~\cite{MAGIC:2016xys} (green solid), as well as the constraints from invisible Higgs decay searches~\cite{Wang:2841498} (gray dashed). The rate factor predicted by the relic density constraint is shown as a function of DM mass (black dashed).}\vspace{-3mm}
\label{fig:HiggsPortal}
\end{figure}

Figure~\ref{fig:HiggsPortal} shows the Higgs portal parameter space with the predicted total annihilation rate as a function of the DM mass. It is intriguing that $\gamma$-ray line searches have the strongest sensitivity around the Higgs resonance, a region in parameter space that is typically challenging to test. We make use of the fact that, given the Fermi-LAT energy resolution, the line signatures can be well distinguished from the continuum emission, such as final state radiation, up to $\gamma$-ray energies of $E_{\gamma \gamma} \sim 300$ GeV, as discussed in Refs.~\cite{Duerr:2015bea,Duerr:2015mva}. 

\section{Implications for the Galactic Center Excess}
Observations of $\gamma$-ray emission from the Milky Way galactic center have long observed a $\gamma$-ray excess that has been named the GCE~\cite{Goodenough:2009gk, Hooper:2011ti, Abazajian:2012pn, Daylan:2014rsa, Calore:2014xka, Fermi-LAT:2017opo}. While the origin of the GCE is disputed, the two most compelling explanations involve DM annihilation or the combined emission from a population of millisecond pulsars (MSPs)~\cite{Hooper:2010mq, Abazajian:2010zy, Hooper:2011ti, Hooper:2013nhl, Cholis:2014lta, Lee:2015fea, Bartels:2015aea, Leane:2019xiy, Leane:2020nmi, Leane:2020pfc, Buschmann:2020adf, Macias:2016nev, Macias:2019omb, Cholis:2019ejx, Cholis:2021rpp}. Within the context of DM models, a DM candidate that annihilates predominantly via $\chi \chi \rightarrow b\bar{b}$ with a mass $m_\chi$ between $40$ GeV and $70$ GeV is highly consistent with the data~\cite{Daylan:2014rsa, Calore:2014xka}.

Since this tree-level final state involves charged particles, there is unavoidably a loop process leading to monoenergetic $\gamma \gamma$ and, as dictated by electroweak symmetry, narrow $Z \gamma$ photon lines. However, the simple $b-$quark loop produces a branching ratio of the order of $\text{BR}_{\gamma \gamma} \sim \alpha_{\rm EM}^2/(4 \pi) \sim 10^{-6}$, which falls far below current experimental sensitivities.

On the other hand, a dominant branching fraction into $b-$quark states in this mass range is hard to explain unless the interaction is related to the quark Yukawa couplings. Thus, we are naturally led to the Higgs-boson mediated scenario. Notably, a Higgs-motivated $b \bar{b}-$annihilation rate that fits the GCE unambiguously predicts bright $\gamma \gamma$ and $Z \gamma$ signals.

\begin{figure}[t!]
\includegraphics[width=\columnwidth]{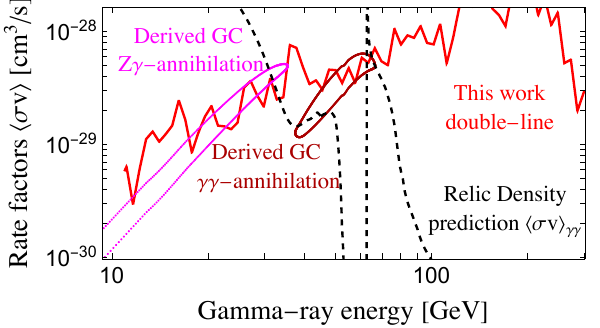}
\caption{ The predictions for the $\gamma \gamma$ (dark red) and $Z \gamma$ (magenta) annihilation rates in the Higgs portal model, given the GCE $95\%$ confidence interval found in Ref.~\cite{Calore:2014xka}. We superimpose our double-line search limits from a gNFW profile in ROI41 (red solid), and the $\langle \sigma v \rangle_{\gamma \gamma}$ values predicted by the freeze-out (black dashed). Note that in contrast to constraints from dwarf galaxies this search for $\gamma-$lines in the GC does not have an intrinsic J-factor uncertainty, when the comparison to the continuum excess is made.}\vspace{-3mm}
\label{fig:DM_GC}
\end{figure}

Figure~\ref{fig:DM_GC} shows the $95\%$ confidence interval for the GCE signal-predictions for the annihilation rates $\langle \sigma v\rangle_{\gamma \gamma}$ and $\langle \sigma v\rangle_{Z \gamma}$, as a red and magenta ellipse, respectively. Those regions are derived from the best-fit region of Ref.~\cite{Calore:2014xka}. We compare the predicted rates with our best limits from the double-line analysis of 15 years of Fermi-LAT data, and find that are results are beginning to be in tension with minimal Higgs portal mediated scenarios as an explanation for the GCE.
By minimal we mean models based on the interactions in Eqs. \ref{eq:ModelScalar}, and \ref{eq:ModelFermion}, in which the relic density is determined by the s-channel Higgs annihilation (shown as black dashed line), and given the searches for invisible Higgs decays apply. Note that other approaches find best-fit parameter regions~\cite{DiMauro:2021qcf}, that are only consistent with the relic density predictions within the excluded parameter range.

\section{Discussion and Conclusion}

In this \emph{letter}, we have reanalyzed 15 years of Fermi-LAT data and studied spectral signatures stemming from DM annihliation to monoenergetic lines. We find no evidence for any statistically significant excesses and set strong limits on DM annihilation to mono-energetic photons. Our results can be applied to a broad range of DM scenarios, constraining DM masses up to (or beyond) the electroweak scale in many well-motivated DM models. If non-perturbative effects, such as bound-state formation are taken into account~\cite{Mitridate:2017izz,Smirnov:2019ngs,Becker:2022iso}, the reach extends to even higher DM masses.

Additionally, we performed the first double-line analysis of the full Fermi-LAT data-set, using 15 years of data. We placed strong limits on thermal DM that is coupled to the Standard Model through the Higgs portal, in a largely model independent way. Our results are in moderate tension (but do not entirely rule out), Higgs Portal models of the GCE with dark matter masses that sit near the m$_H$/2 resonance. This parameter space is of interest due to the fact that it is the only portion of the Higgs Portal parameter space that is consistent with the GCE and constraints from the branching ratios to invisible particles. Moreover, we note that our constraints are roughly independent of the dark matter density profile near the galactic center, as the cross-sections for both the GCE continuum and the line search shift in the same way.

We emphasise, that the double-line technique can be applied to other datasets, and is particularly promising at energies near the Higgs resonance. Furthermore, it has an enhanced discovery potential for $\gamma$-ray signals that have a limited sensitivity due to low photon counts, as it makes use of additional information from photons in correlated energy bins.

\section*{Acknowledgements}
We would like to thank Ben Safdi, Josh Foster, Rebecca Leane, and Linda Xu for helpful comments and discussions. PD and TL are supported in part by the European Research Council under grant 742104 and the Swedish National Space Agency under contract 117/19. JS was also supported  by the European Research Council under grant 742104 during the initial stage of the project. TL is also supported by the Swedish Research Council under contracts 2019-05135 and 2022-04283. This project used computing resources from the Swedish National Infrastructure for Computing (SNIC) under project Nos. 2021/3-42, 2021/6-326, 2021-1-24 and 2022/3-27 partially funded by the Swedish Research Council through grant no. 2018-05973.


\clearpage
\newpage
\onecolumngrid

\appendix

\section{Fermi counts spectra and signal fit}
\label{sec:AppendixA}

In this appendix we show examples of the derived count spectra, and the template functions used in the single- and double-line analyses in the first figure panel. 
We furthermore show the limits obtained from the single-line analysis for all ROIs studied in this work in the second figure panel, and for the double-line analysis in the third figure panel. Finally, we report the ratio BR$_{Z \gamma}/$BR$_{\gamma\gamma}$ as function of the DM mass, as derived from~\cite{LHCHiggsCrossSectionWorkingGroup:2011wcg} (consistent with analytic calculations from~\cite{Djouadi:2005gi}), which motivates the application of the double-line analysis for $\mathcal{O}$(TeV) gamma-ray data

\begin{figure*}[!h]
\centering
\includegraphics[width=0.45\textwidth] {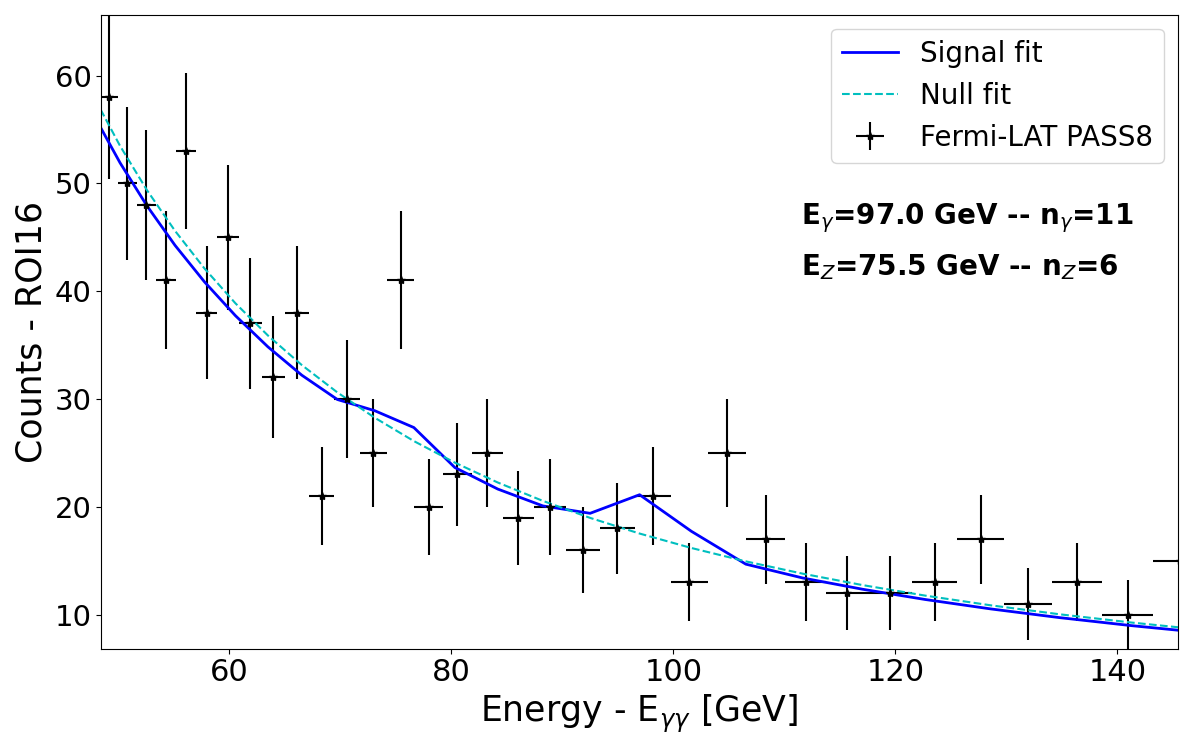}
\hspace{0.5cm}
\includegraphics[width=0.45\textwidth] {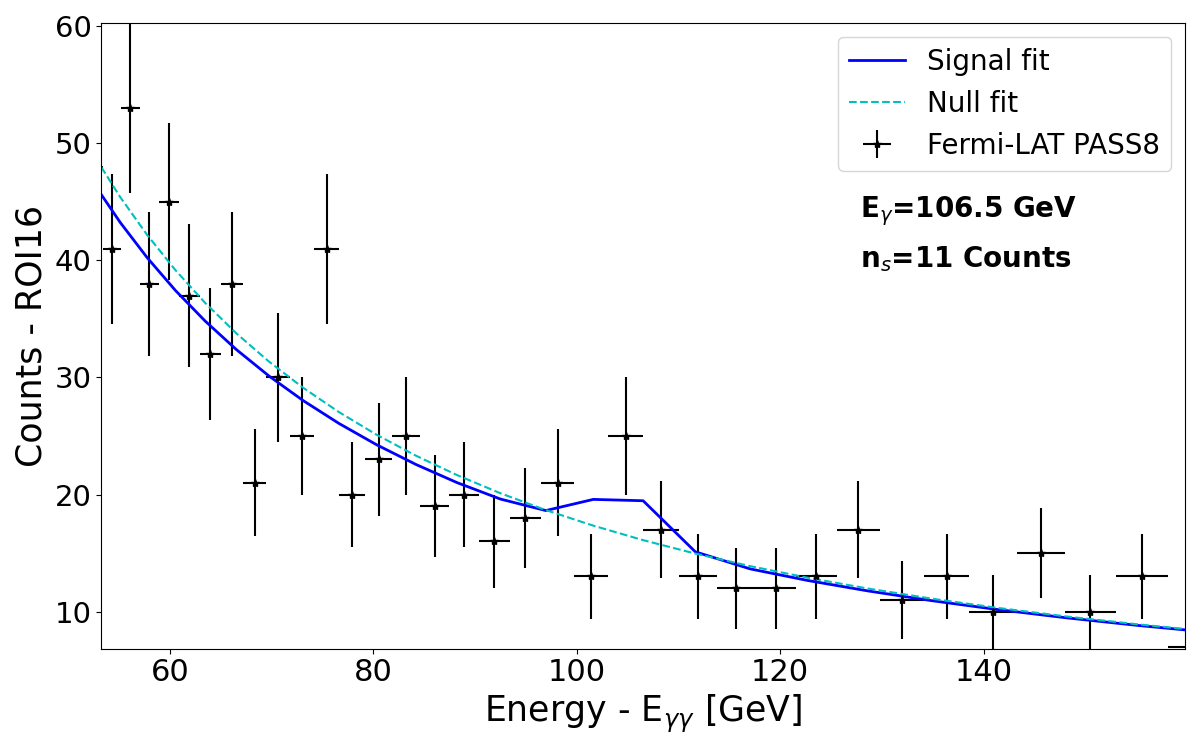}
\includegraphics[width=0.45\textwidth] {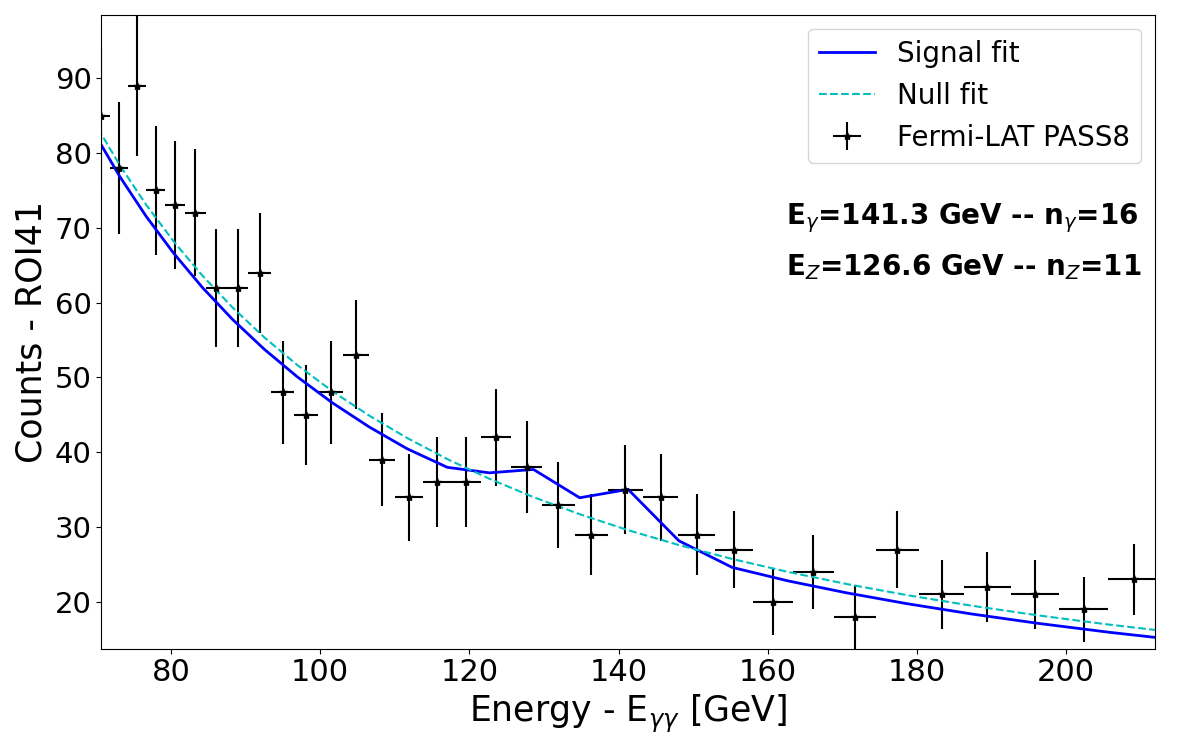} 
\hspace{0.5cm}
\includegraphics[width=0.45\textwidth] {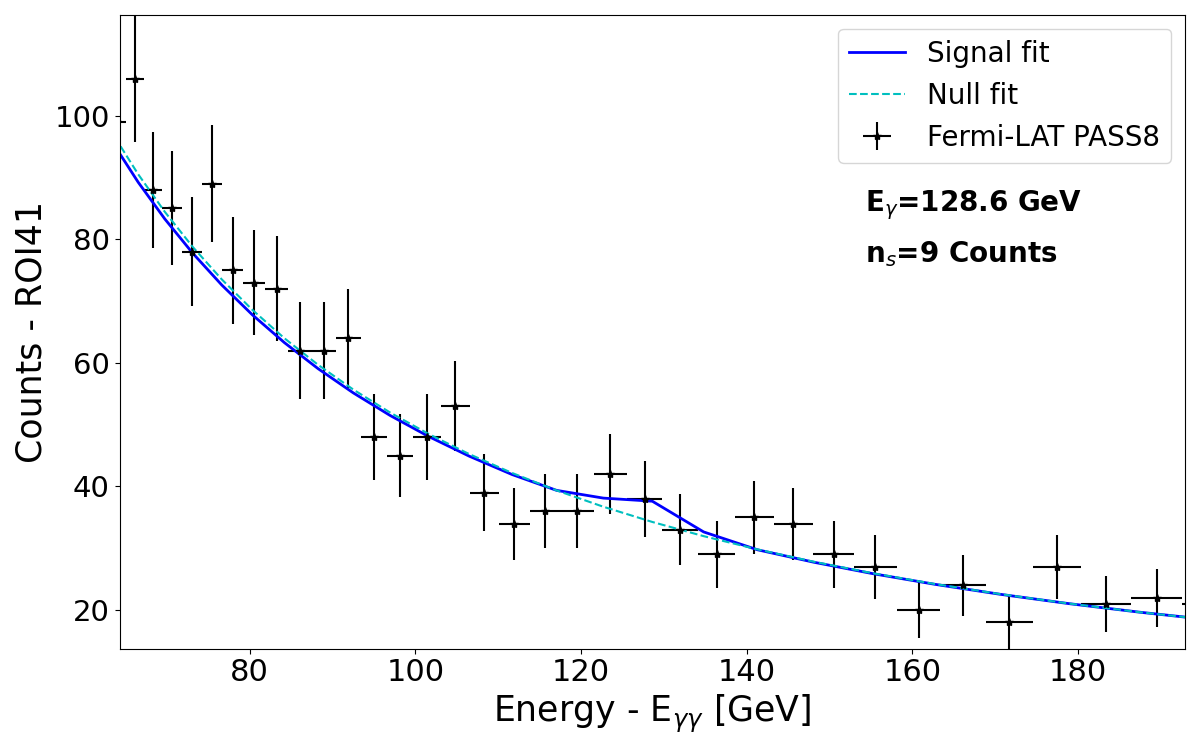} 
\caption{Count spectra for two different energy windows comparing the double-line (left column) and single-line (right column) best-fit signals. In the upper panels, the energy window is centered at E$_{\gamma\gamma}\sim106$~GeV (E$_{Z\gamma}\sim87$~GeV) and in the lower panels at E$_{\gamma\gamma}\sim155$~GeV (E$_{Z\gamma}\sim142$~GeV). At energies greater than E$_{\gamma\gamma}\sim180$~GeV both signals expected in the double-line analysis merge.}
\label{fig:Counts}
\end{figure*}

Figure~\ref{fig:Counts} shows the count spectrum for the ROI16 region at two different energies: at around $106$~GeV (upper row), where fluctuations at the level of $1-2\sigma$ are observed for all ROIs (see Fig.~\ref{fig:TS_comp}) and at $155$~GeV, where the signals from the photons produced in the $\gamma \gamma$ and $Z\gamma$ decay start to merge into the same energy bin (leading to a significantly stronger constraint compared to the one from the single-line analysis above this energy). Here, we include the fitted count spectra assuming only background (null fit) and assuming background+signal (signal fit), which allows us to see a comparison of the signals searched in the single-line (right panels) and double-line (left panels) analysis.

\newpage

\section{Single-line limits}
\label{sec:AppendixB}

\begin{figure*}[ht]
\centering
 \includegraphics[width=0.45\textwidth] {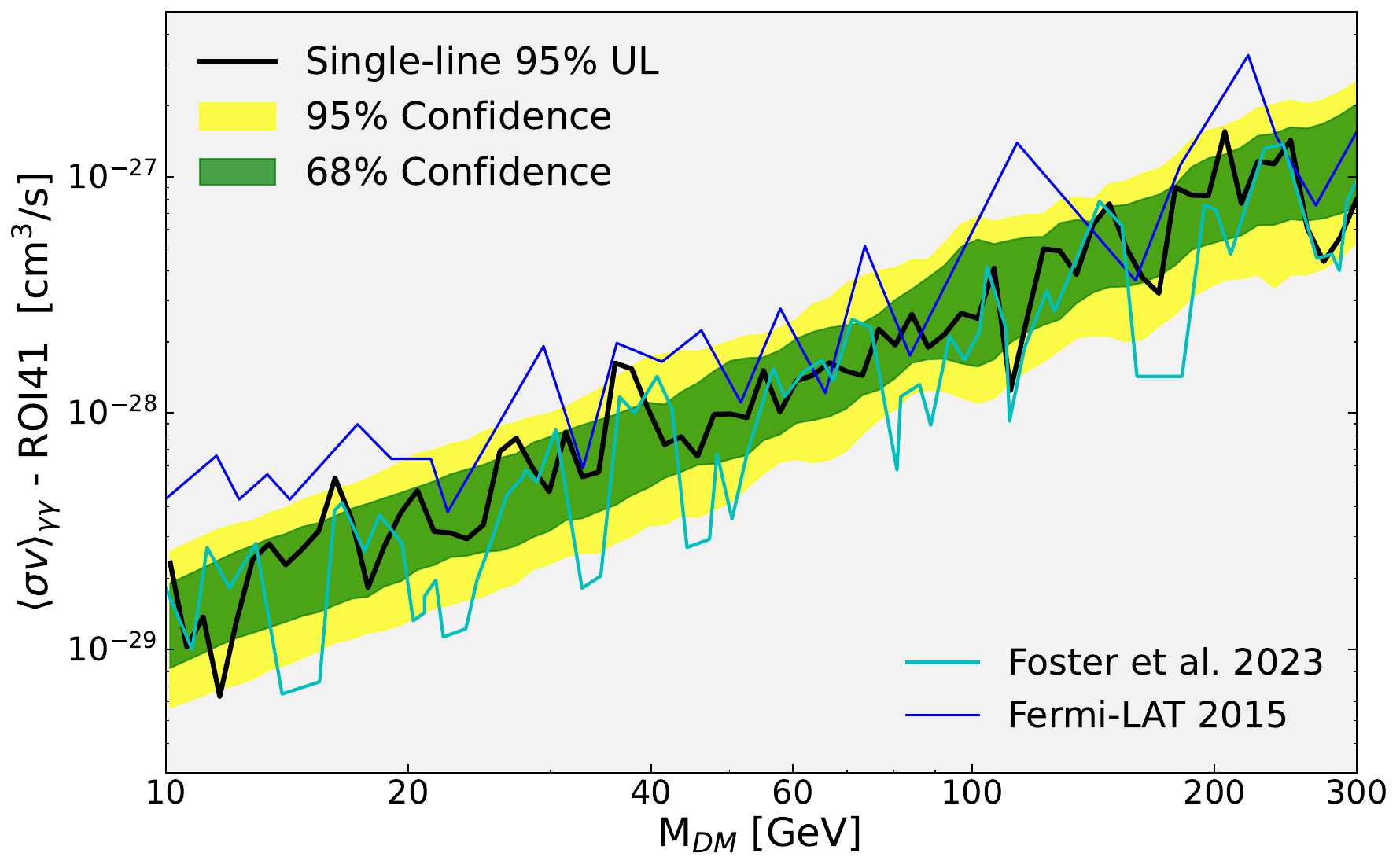}  \hspace{0.4cm}
\includegraphics[width=0.45\textwidth] {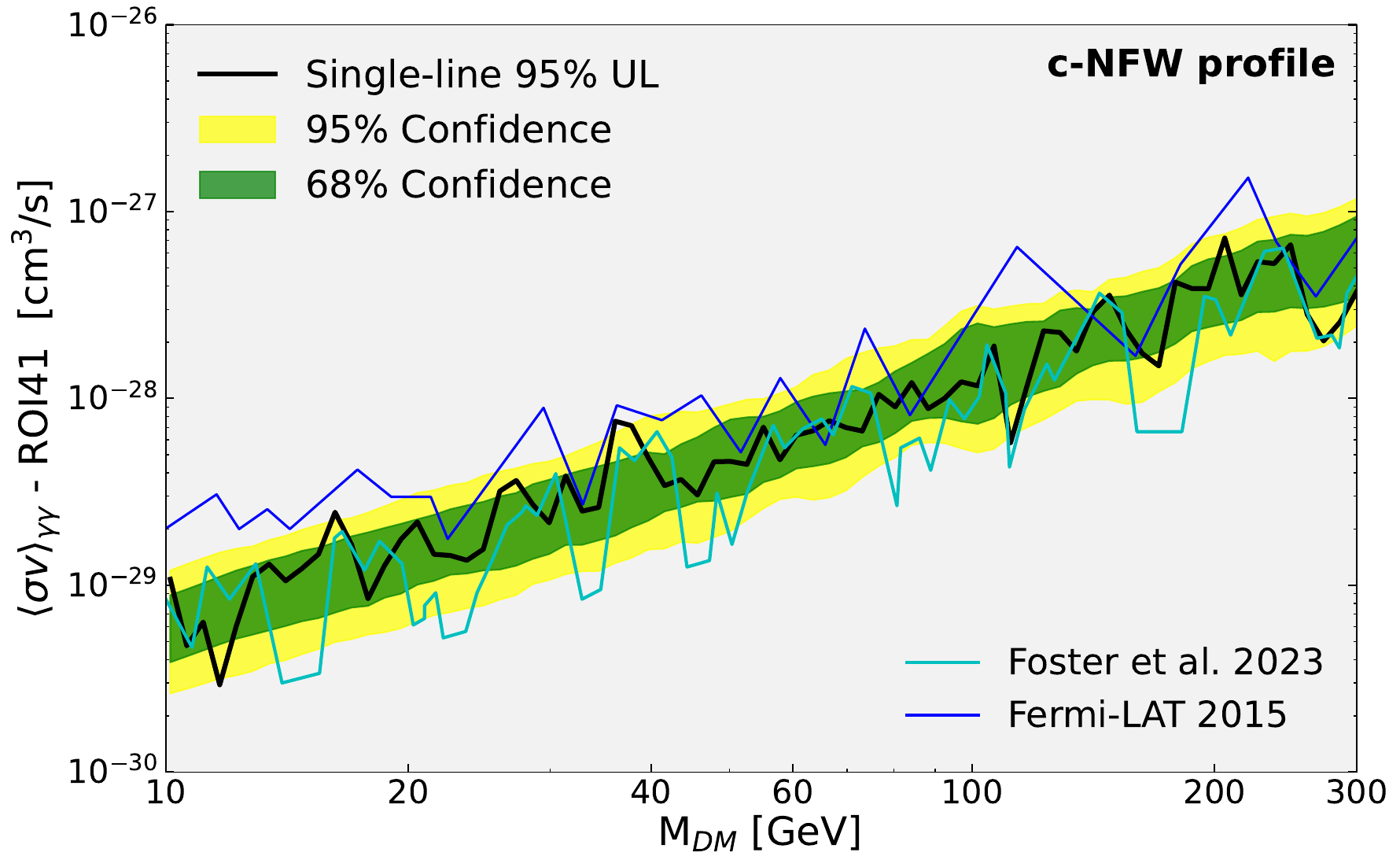} 
\includegraphics[width=0.45\textwidth] {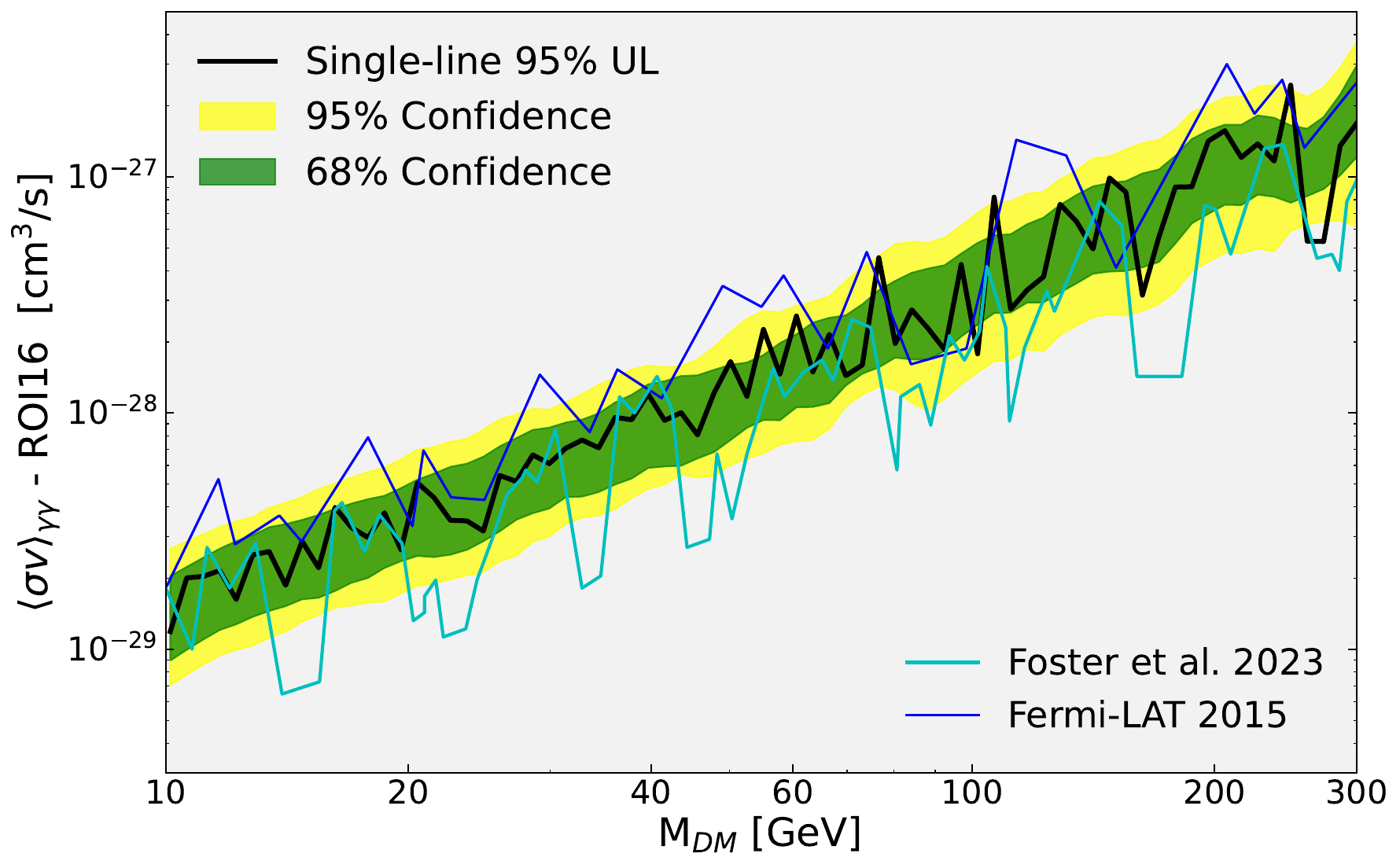} \hspace{0.4cm}
\includegraphics[width=0.45\textwidth]
{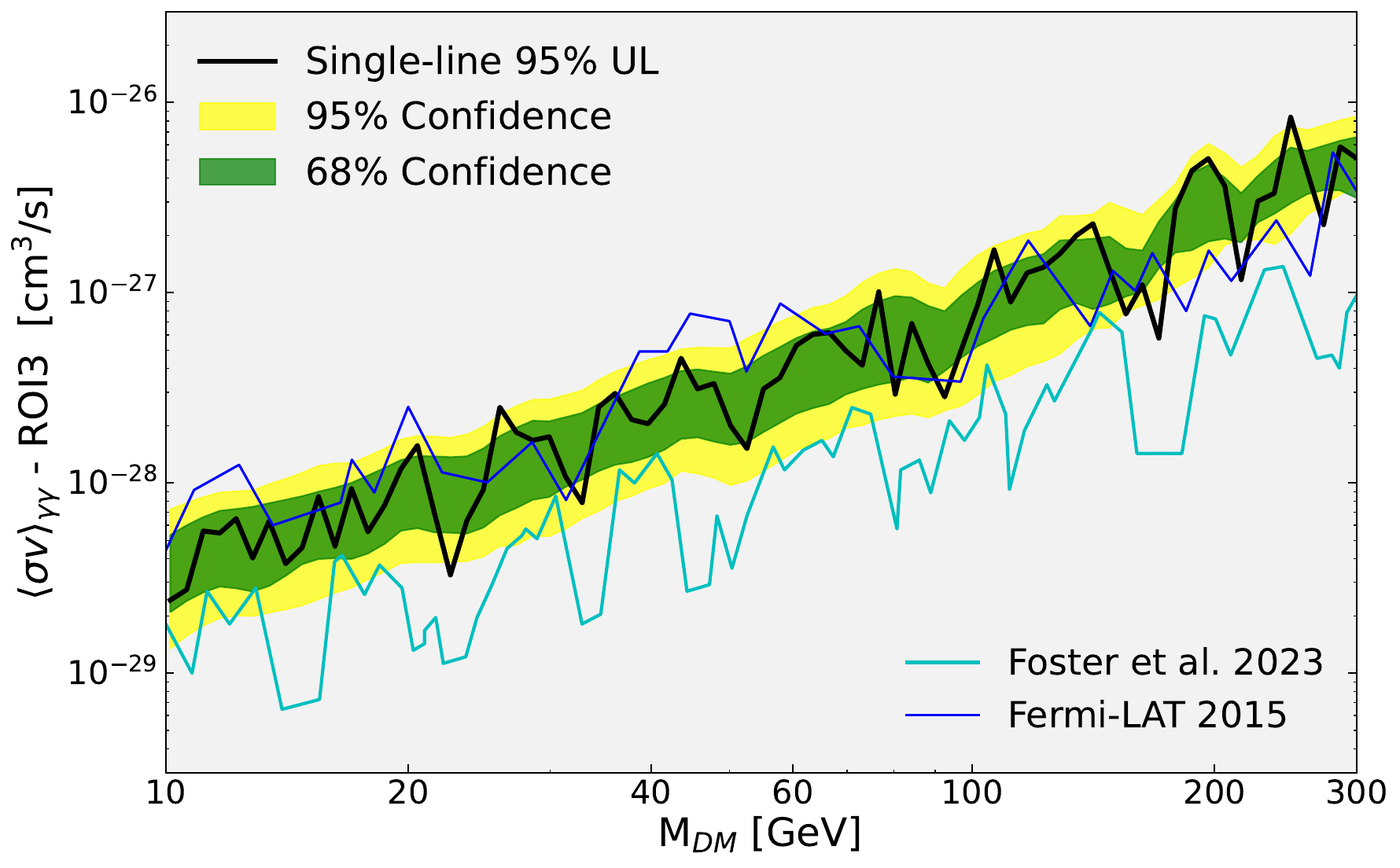}
\caption{DM bounds and confidence bands for different ROIs obtained from the single-line analysis, compared to those from Refs.~\cite{DAMPE_Lines, Fermi2015, Foster}. The upper panels show the limits for the ROI16 region, assuming an NFW and a contracted-NFW (with index $\gamma=1.3$) DM profile, for the upper-left and upper-right panels, respectively. The limits from other works have been rescaled in the case where we show the limits assuming a c-NFW profile. The lower panels show the limits for the ROI3 (left) and ROI41 (right) regions.}
\label{fig:DM_limits_SL}
\end{figure*}

Figure~\ref{fig:DM_limits_SL} shows the limits obtained from the single-line analysis compared to those from Refs.~\cite{DAMPE_Lines, Fermi2015, Foster}. The ROI16 region constitutes the region that leads to better bounds. Thus, in the upper panels, we show the limits derived assuming an NFW profile, on the left, and a contracted-NFW profile (with index $\gamma=1.3$), on the right. The contracted-NFW profile has been found in several papers to be the DM profile most compatible with the GCE. In this case, the limits from other works have been rescaled accordingly. Then, in the lower panels we show the limits obtained for the ROI3 (left) and ROI41 (right) regions. As we see, the ROI3 region offers the most conservative limits, which is due to the smaller ROI, which produces many fewer counts and larger Poissonian uncertainties, while the ROI41 region leads to similar but slightly higher bounds than in the ROI16 region. Larger ROIs can have large systematic uncertainties associated to the analysis and a lower expected signal-to-noise ratio for a DM signal.

\newpage

\section{Double-line limits}
\label{sec:AppendixB}
\begin{figure*}[hb]
\centering
\includegraphics[width=0.49\textwidth] {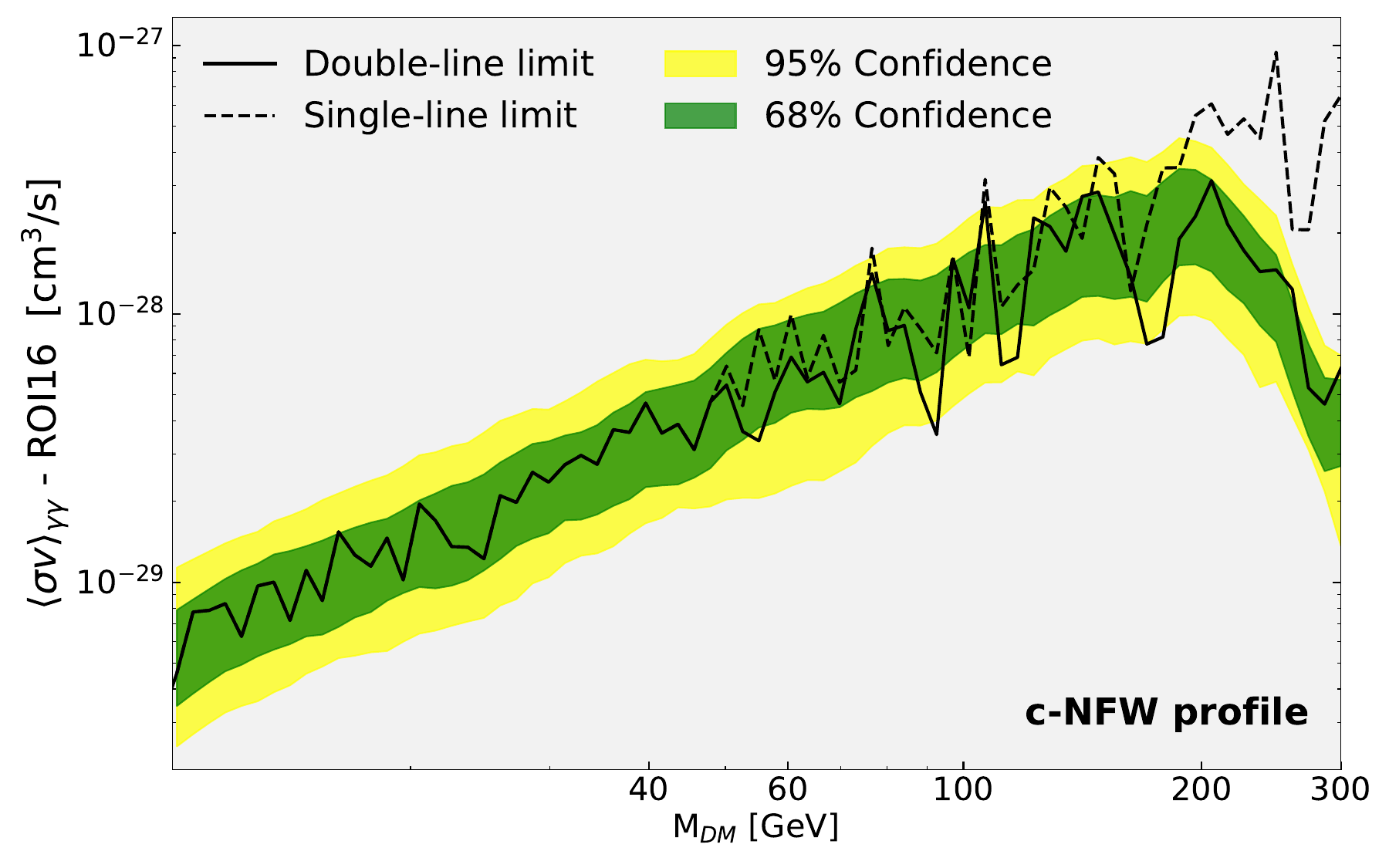} 
\includegraphics[width=0.49\textwidth]
{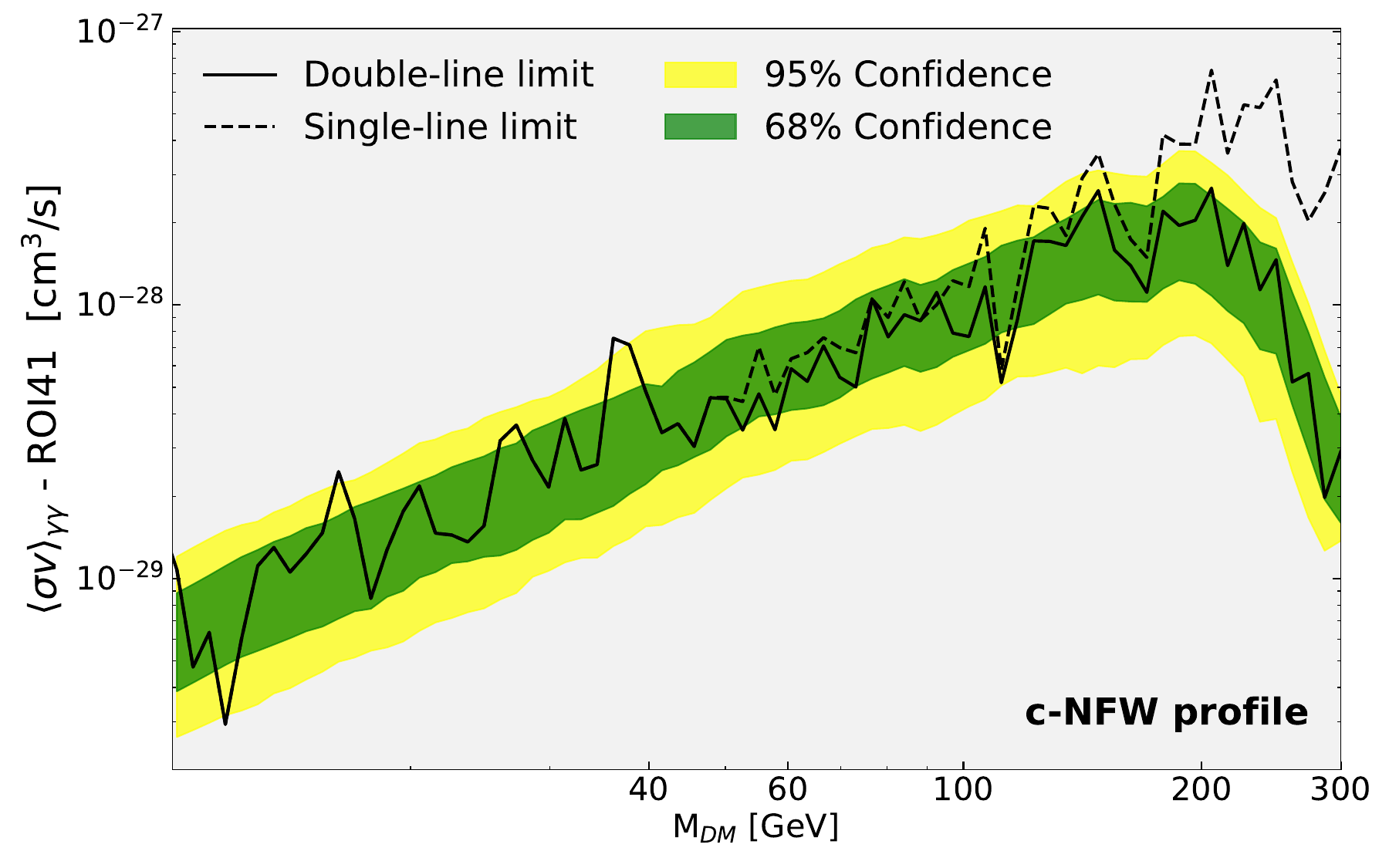}
\caption{DM bounds and confidence bands for different ROIs obtained from the double-line analysis, compared to the limits derived from the single-line analyses, for the ROI16 (left panel) and ROI41 (right panel), assuming a c-NFW profile (with index $\gamma=1.3$).}
\label{fig:DM_limits_DL}
\end{figure*}

Figure~\ref{fig:DM_limits_DL} shows the DM bounds and confidence bands obtained from the double-line analysis for the ROI16 (left panel) and ROI41 (right panel), compared to the limits derived from the single-line analyses, drawn assuming a c-NFW profile (with index $\gamma=1.3$). This show that the double-line analysis would constitute a more sensitive way to look for line-like gamma-ray signals at very high energies. The reason is that at high energies the photons from the $\gamma \gamma$ and $Z\gamma$ decays are produced at roughly the same energy and the branching ratio for the $Z\gamma$ decay becomes much higher than for the $\gamma \gamma$ process, leading to a more constraining limit of $\langle \sigma v\rangle_{\gamma \gamma}$. This means that a search for lines at the energies where the $Z\gamma$ decay dominates lead to a much stronger constraint due to the sum of the signals coming from the $\gamma \gamma$ and $Z\gamma$ decays and the branching ratio for the $Z\gamma$ process. 

\begin{figure}[th!]
\includegraphics[width= 0.5\columnwidth]{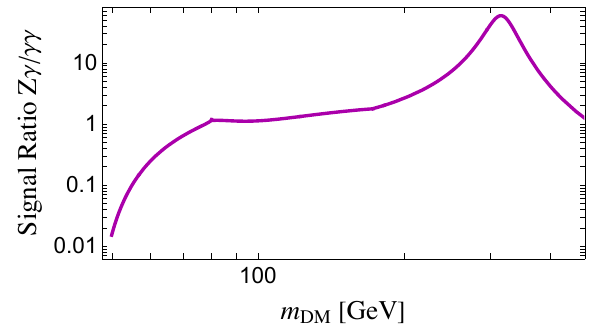}
\caption{The signal-strength ratio of the $Z \gamma$ and $\gamma \gamma$ mono-energetic photon signals for Higgs mediated annihilation~\cite{LHCHiggsCrossSectionWorkingGroup:2011wcg}.}\vspace{-3mm}
\label{fig:ZGamma}
\end{figure}

Figure~\ref{fig:ZGamma} displays the signal strength ratio between the  $Z \gamma$ and $\gamma \gamma$ mono-energetic photon signals for Higgs mediated annihilation. At low energies the massive Z-boson emission suppresses the signal, in the intermediate mass range the ratio is determined by the coupling ratio between the weak and electromagnetic couplings $\alpha_{\rm EW}\sim 1/29$ and $\alpha_{\rm EM} \sim 1/137$, as well as the photon multiplicity. Finally, at larger masses the loop factor of the $\gamma \gamma$ signal is suppressed due to the destructive interference of virtual particles, which enhances the relative $Z\gamma$ signal strength.

\clearpage

\twocolumngrid
\bibliography{main}

\end{document}